\documentclass[aps,pra,nofootinbib,notitlepage,twocolumn]{revtex4-1}

\usepackage{amsmath,amssymb,amsfonts,graphicx}
\usepackage{color}
\usepackage{subfigure}
\usepackage[hidelinks]{hyperref}

\renewcommand{\vec}[1]{{\mathbf{#1}}}
\newcommand{\beq}{\begin{eqnarray}}
\newcommand{\eeq}{\end{eqnarray}}
\newcommand{\tabi}{\hspace{.1\textwidth}}

\begin{document}

\title{Violation of f-sum Rule with Generalized Kinetic Energy}
\author{Kridsanaphong Limtragool and Philip W. Phillips}

\affiliation{Department of Physics and Institute for Condensed Matter Theory,
University of Illinois
1110 W. Green Street, Urbana, IL 61801, U.S.A.}
\date{\today}

\begin{abstract}
Motivated by the normal state of the cuprates in which the f-sum rule increases faster than a linear function of the particle density, we derive a conductivity sum rule for a system in which the kinetic energy operator in the Hamiltonian is a general function of the momentum squared. Such a kinetic energy arises in scale invariant theories and can be derived within the context of holography.  Our derivation of the f-sum rule is based on the gauge couplings of a non-local Lagrangian in which the kinetic operator is a fractional Laplacian of order $\alpha$. We find that the f-sum rule in this case deviates from the standard linear dependence on the particle density.  We find two regimes.  At high temperatures and low densities, the sum rule is proportional to $nT^{\frac{\alpha-1}{\alpha}}$ where $T$ is the temperature. At low temperatures and high densities, the sum rule is proportional to $n^{1+\frac{2(\alpha-1)}{d}}$ with $d$ being the number of spatial dimensions. The result in the low temperature and high density limit, when $\alpha < 1$, can be used to qualitatively explain the behavior of the effective number of charge carriers in the cuprates at various doping concentrations.

\end{abstract}

\maketitle

\section{Introduction}
Understanding the nature of the current carrying degrees of freedom in the normal states of the superconducting copper oxides stands as a key challenge in modern condensed matter physics. Many properties in the normal states of the cuprates deviate from the standard theory of metals. One well-known example is that the electrical resistivity, $\rho$, observed in the normal state, exhibits a non-Fermi liquid behavior. Instead of having $\rho \propto T^2$ as in the case of Fermi liquid, $\rho$ in the cuprates goes like $T^a$ with $a$ in a range of $1$ to $2$ depending on the chemical composition\cite{Naqib2003}. Explaining such strange properties in the cuprates may require a non-traditional model, in particular models in which the basic notions of particles and locality are abandoned.

\begin{figure}[h!]
	\includegraphics[scale=0.5]{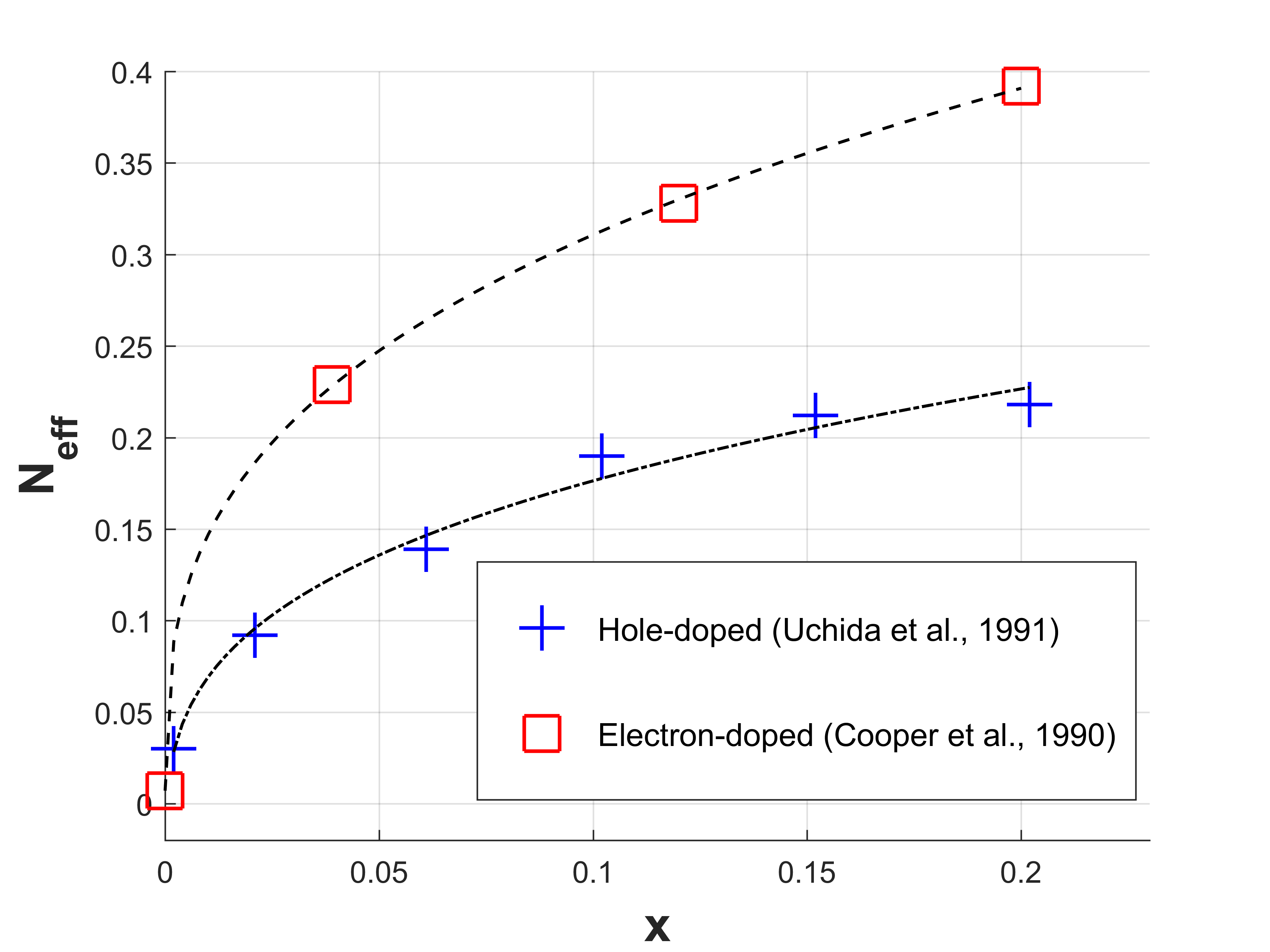} 
    \caption{Effective number of charge carriers ($N_{\rm eff}$) vs. doping concentration ($x$) from Refs. \cite{Cooper1990,Uchida1991}. } \label{fig:neffvsx}
\end{figure}

The focus of this study is  the deviation of the integrated spectral weight of the optical conductivity (also known as an optical sum) in the normal states of the cuprates from the standard f-sum rule (or conductivity sum rule). The content of the f-sum rule is that the optical sum is directly proportional to the charge carrier density: $\int_{0}^\infty \sigma_1(\omega)\omega = \frac{\pi e^2 n}{2 m}$. Here $\sigma_1$ is the real part of the optical conductivity, $n$ is the charge carrier density, $e$ is the electric charge, and $m$ is the mass. When $\sigma_1(\omega)$ is integrated up to a cutoff frequency $\omega_c$, the optical sum is proportional to the effective number of charges from energy below $\omega_c$ ($N_{\rm eff}$). In normal metals, when $\omega_c$ is chosen to be in the region of the optical gap, $N_{\rm eff}$ is simply given by the number of electrons in the conduction band. However, in the cuprates\cite{Cooper1990,Uchida1991}, $N_{\rm eff}$ deviates from what one expects from the dopant concentration, $x$. When $0<x<0.2$, instead of having $N_{\rm eff}(x) = x$, $N_{\rm eff}(x)$ is greater than $x$ and is concave downward. We find that the empirical $N_{\rm eff}$ from Refs. \cite{Cooper1990,Uchida1991} can be fitted to the functional form,
\beq \label{eq:neffx}
N_{\rm eff} = N_{0} + N_1 x^\gamma,
\eeq
with $\gamma \approx 0.3 - 0.4$\footnote{We fitted Eq. (\ref{eq:neffx}) to the data points extracted from the plots in Ref. \cite{Cooper1990,Uchida1991}. As a result, the values of $\gamma$ we present here are only approximated.}. Here $N_0$ and $N_1$ are dimensionless constants. Shown in Fig. \ref{fig:neffvsx} are the plots of $N_{\rm eff}$ as a function of $x$ from Refs. \cite{Cooper1990,Uchida1991} overlaid with the fitted lines from Eq. (\ref{eq:neffx}).

The proof (see for example \cite{Kubo1957,Pines1999,PinesNozieres1999}) underlying the conductivity sum rule relies on the fact that the kinetic energy operator of a single particle in the Hamiltonian is $K = \frac{p^2}{2m}$. The deviation from the standard sum rule indicates that the dynamics of the charge carrying degrees of freedom may not be governed by the kinetic term which is quadratic in momentum. Recently, in the context of the gauge/gravity duality, one of us\cite{LaNave2016} has shown that a massive free theory with a geodesically complete metric in the bulk generically gives rise to a boundary theory with a fractional Riesz derivative (a fractional Laplacian). The power of the fractional derivative is partially determined by the mass of the field. The result of this work implies that, in some cases, the infrared behavior of a strongly coupled theory could be described by a non-local operator such as a fractional derivative. This leads us to a postulate that an emergent charge carrier in the infrared is an object with a fractional kinetic energy.  That is, the kinetic energy operator is a fractional Riesz derivative $K \propto (-\partial^2)^\alpha$ with $\alpha$ being a positive real number. Equivalently, in momentum space, the kinetic term is a fractional power of momentum $K\propto p^{2\alpha}$. We note that the quantum mechanics of such a kinetic operator was studied in Refs. \cite{Laskin2000,Laskin2000a,Laskin2002}. Recently, the fractional kinetic operator has been presented as a way of understanding unparticles\cite{Domokos2015}.

In this work, we consider a model of non-relativistic particles with a kinetic term given by a general function of momentum squared, $K(\vec p^2)$. The particles are allowed to have non-derivative interactions with one another. This model is equivalent to the restricted band model where the kinetic energy is replaced by the band dispersion, $E(\vec p)$.\footnote{We ignore the fact that the kinetic energy of our model is rotationally invariant and simply replace it by the band dispersion.} In the restricted band model, one considers only particles in a single band and ignores the inter-band interactions. It turns out that the conductivity sum rule of the restricted band model\cite{Kubo1957,Benfatto2006}, is given by
\beq \label{eq:sumrule_restrictedband}
W \equiv \int\limits_{0}^{\infty} \sigma_{1}(\omega) d\omega &=& \frac{\pi e^2}{2} \int \frac{d^dp}{(2\pi)^d} n(\vec p)\frac{\partial^2}{\partial p_i^2}E(\vec p), \nonumber \\
\eeq
where $\sigma_1$ is the real part of the optical conductivity and $n(\vec p)$ is the occupation number of the momentum state $\vec p$. We review a proof of the sum rule in this paper. Our proof is based on the gauge couplings of a nonlocal Lagrangian\cite{Terning1991}. This sum rule is applied in many systems such as the Hubbard model\footnote{The sum rule in this case is usually written as \beq
W \equiv \int\limits_{0}^{\infty} \sigma_{1}(\omega) d\omega &=& -\frac{\pi e^2}{2}a_i^2\langle K_i \rangle \nonumber
\eeq
where $a_i$ and $K_i$ are the lattice spacing and the kinetic energy operator along the $i$th direction, respectively.}\cite{Maldague1977,Baeriswyl1987}, graphene\cite{Gusynin2007}, and the d-density wave state\cite{Benfatto2005,Benfatto2006}. We then apply the conductivity sum rule to the case of non-interacting fermions with fractional kinetic energy: $K(\vec p^2) \propto p^{2\alpha}$. We show that the behavior can be divided in two regimes. In the high temperature and low density regime, the sum rule is proportional to $n T^{\frac{\alpha-1}{\alpha}}$ where $n$ is the density and $T$ is the temperature. On the other hand, in the low temperature and high density regime, the sum rule is proportional to $n^{1 + \frac{2(\alpha-1)}{d}}$. Here $d$ denotes the number of spatial dimensions. To make contact with experiment, we make a further assumption that the density of these emergent excitations, $n$, is the same as the density of bare charge carrier (bare electrons or holes). This means $n \propto x$ in the cuprates. In the low temperature and high density limit with $0 < \alpha < 1$, the optical sum is proportional to $x^\beta$ with $0 < \beta =  1 + \frac{2(\alpha-1)}{d} < 1$ which is qualitatively the same behavior as $N_{\rm eff}$ in the cuprates.

\section{Hamiltonian with a Generalized Kinetic Energy}
We investigate a system of non-relativistic particles in which its kinetic term has a non-canonical form. $K$ is not necessarily proportional to a square of momentum ($\vec p^2$) but is some general function of $\vec p^2$, i.e. $K = K(\vec p^2)$. The second quantized Hamiltonian of this system in $d$ spatial dimensions is
\beq \label{eq:H}
H = \int d^dr \psi^\dagger(\vec r)\bigg[K(-\partial^2)-\mu \bigg]\psi(\vec r) + H_{\rm other},
\eeq
where $\psi^\dagger(\vec r)$ and $\psi(\vec r)$ are creation and annihilation field operators, respectively, $\mu$ is the chemical potential, and $H_{other}$ describes non-derivative potentials and interactions. Since $H_{\rm other}$ contains no derivative operators, the current only comes from the kinetic term. To derive the conductivity sum rule of this model, one needs the form of its $U(1)$ current operator. 

\subsection{Current Operator}
The couplings between the particle fields and the $U(1)$ electromagnetic gauge fields can be obtained by gauging a nonlocal Lagrangian with Wilson lines\cite{Mandelstam1962,Terning1991}. One starts by rewriting the kinetic term of the Hamiltonian, $H_K =  \int d^dr \psi^\dagger(\vec r)K(-\partial^2)\psi(\vec r) $, in the form,
\beq
H_K = \int d^drd^dr' \psi^\dagger(\vec r)F(\vec r,\vec r')\psi(\vec r'),
\eeq
where $F(\vec r,\vec r')$ is a function resulting from rewriting the kinetic term. $H_K$ can be made $U(1)$ invariant by inserting a Wilson line, $W(\vec r,\vec r') = \exp(-ie\int_{\vec r}^{\vec r'} dx^i A_i(\vec{x}))$, between $\psi^\dagger(\vec r')$ and $\psi(\vec r)$ in the kinetic term as
\beq
H_K = \int d^drd^dr' \psi^\dagger(\vec r)W(\vec r,\vec r')F(\vec r,\vec r')\psi(\vec r').
\eeq
Here $e$ is the electric charge and $A_i$ is the $i$th component of a $U(1)$ electromagnetic gauge field. The vertex couplings can be derived by taking derivatives of the gauged $H_K$ with respect to the particle and gauge fields. The coupling between two particles and one gauge field is
\beq
e\Gamma^i(\vec p,\vec q) &=& \frac{\delta^3 H_K}{\delta A_i(\vec q)\delta\psi(\vec p)\delta\psi^\dagger(\vec p + \vec q)} \nonumber \\ 
&=& e(2\vec p+\vec q)^i\mathcal{F}(\vec p,\vec q)
\eeq
and the coupling between two particles and two gauge fields is
\begin{widetext}
\beq
e^2\Gamma^{ij}(\vec p,\vec q_1,\vec q_2) &=& \frac{\delta^4 H_K}{\delta A_i(\vec q_1)\delta A_j(\vec q_2)\delta\psi(\vec p)\psi^\dagger(\vec p + \vec q_1 + \vec q_2)} \nonumber \\
&=& e^2\bigg\{ 2\delta^{ij}\mathcal{F}(\vec p,\vec q_1+\vec q_2)  +\frac{(2\vec p+\vec q_2)^j(2\vec p+2\vec q_2+\vec q_1)^i}{\vec q_1^2+2(\vec p+\vec q_2)\cdot \vec q_1}[\mathcal{F}(\vec p,\vec q_1+\vec q_2)-\mathcal{F}(\vec p,\vec q_2)] \nonumber \\
&& +\frac{(2\vec p+\vec q_1)^i(2\vec p+2\vec q_1+\vec q_2)^j}{\vec q_2^2+2(\vec p+\vec q_1)\cdot \vec q_2}[\mathcal{F}(\vec p,\vec q_1+\vec q_2)-\mathcal{F}(\vec p,\vec q_1)]\bigg\},
\eeq
\end{widetext}
with 
\beq
\mathcal{F}(\vec p,\vec q) = \frac{K((\vec p+\vec q)^2) - K(\vec p^2)}{(\vec p+\vec q)^2-\vec p^2}.
\eeq
Using the vertex couplings obtained above, one can expand $H_K$ to second order in gauge fields as
\begin{align} \label{eq:HA}
H_K = &\int d^dr \psi^\dagger(\vec r)K(-\partial^2)\psi(\vec r) \nonumber \\
& + e\int\frac{d^dpd^dq}{(2\pi)^{2d}}\psi^{\dagger}(\vec p+ \vec q)\psi(\vec p)\Gamma^i(\vec p,\vec q)A_i(\vec q) \nonumber \\
& + \frac{e^2}{2}\int\frac{d^dpd^dq_1d^dq_2}{(2\pi)^{3d}}\psi^{\dagger}(\vec p+\vec q_1+\vec q_2)\psi(\vec p) \nonumber \\
& \tabi \ \ \ \ \ \ \times\Gamma^{ij}(\vec p,\vec q_1,\vec q_2)A_i(\vec q_1)A_j(\vec q_2)\nonumber \\
& + O(A^3).
\end{align}
We neglect the higher order terms, since we only need up to the terms with two gauge fields in  linear response theory. The current operator can be obtained by taking derivatives of $H_K[A_i]$ with respect to the gauge field, 
\beq
J_i(-\vec q) = -(2\pi)^d \frac{\delta H_K}{\delta A_i(\vec q)}. 
\eeq
Performing the derivative leads to
\begin{align} \label{eq:J_i}
J_i(\vec q) =& -e\int\frac{d^dp}{(2\pi)^{d}}\psi^{\dagger}(\vec p-\vec q)\psi(\vec p)\Gamma^i(\vec p,-\vec q) \nonumber \\
& - e^2\int\frac{d^dp_1d^dp_2}{(2\pi)^{2d}}\psi^{\dagger}(\vec p_1+\vec p_2-\vec q)\psi(\vec p_1) \nonumber \\
& \tabi \times\Gamma^{ij}(\vec p_1,-\vec q,\vec p_2)A_j(\vec p_2).
\end{align}

\section{Derivation of an Optical Sum Rule}
We use linear response theory to derive the conductivity sum rule. Our approach is based on the derivation of the standard conductivity sum rule from Ref. \cite{Millis2004}. The idea on the diamagnetic contribution to the conductivity and some of the notations we use are from Ref. \cite{Tong}. We assume that the system is time-translationally invariant and the background electric field is uniform. We work in the gauge that $A_0 = 0$. Let us denote $\langle O \rangle$ as an expectation value of an operator $O$ with respect to the thermal equilibrium state in the presence of a background gauge field $A_i$. $\langle O \rangle_0$ denotes a thermal expectation value of an operator $O$ with $A_i = 0$. From linear response theory\cite{Giuliani2005}, the difference in the current $\delta \langle J_i(\vec x,t) \rangle \equiv \langle J_i \rangle - \langle J_i \rangle_0$  is given by
\beq \label{eq:lin_response}
\delta \langle J(\vec x,t) \rangle = -i \int\limits_{-\infty}^{t} dt' \int d^dx' \langle [J_i(\vec x,t),J_j(\vec x',t')] \rangle_0 A_j(\vec x',t'). \nonumber \\
\eeq
The total current is then $\langle J_i \rangle = \langle J_i \rangle_0 + \delta \langle J_i \rangle $. The term $\langle J_i \rangle_0$ gives rise to the diamagnetic conductivity, $\sigma^d$, while the term $\delta \langle J_i \rangle$ contributes to the paramagnetic conductivity, $\sigma^p$. Let us first calculate the diamagnetic conductivity. Taking the expectation value, $\langle...\rangle_0$ of Eq. (\ref{eq:J_i}), one has
\begin{align} \label{eq:J_i_expect0}
\langle J_i \rangle_0 (\vec q,\omega) =& -e\int\frac{d^dp}{(2\pi)^{d}}\langle \psi^{\dagger}(\vec p-\vec q)\psi(\vec p) \rangle_0 \Gamma^i(\vec p,-\vec q) \nonumber \\
& - e^2\int\frac{d^dp_1d^dp_2}{(2\pi)^{2d}} \langle  \psi^{\dagger}(\vec p_1+\vec p_2-\vec q)\psi(\vec p_1) \rangle_0 \nonumber \\
& \tabi \ \ \ \times\Gamma^{ij}(\vec p_1,-\vec q,\vec p_2)A_j(\vec p_2,\omega,).
\end{align}
We drop the first term because in the thermodynamic limit ($\vec q \rightarrow 0$), it corresponds to a spontaneous current which vanishes according to the Bloch theorem (see Appendix \ref{app:bloch}). For a uniform background field, we have $A(\vec p_2,\omega) = (2\pi)^d\delta(\vec p_2)A(\omega)$. Integrating over the delta function, $\langle J_i \rangle_0$ can be simplified to
\begin{align}
\langle J_i \rangle_0 (\vec q,\omega) =& -e^2\int\frac{d^dp_1}{(2\pi)^{d}} \langle  \psi^{\dagger}(\vec p_1- \vec q)\psi(\vec p_1) \rangle_0 \nonumber \\
& \tabi \times\Gamma^{ij}(\vec p_1,-\vec q,0)A_j(\omega,).
\end{align}
From the definition of an electrical conductivity $\langle J_i\rangle_0(\vec q,\omega)  = \sigma_{ij}(\vec q,\omega) E_{j}(\vec q,\omega)$, we can extract the diamagnetic conductivity as
\begin{align}
\sigma^d_{ij}(\vec q,\omega) =& \frac{ie^2}{\omega+i\eta}\int\frac{d^dp_1}{(2\pi)^{d}} \langle  \psi^{\dagger}(\vec p_1- \vec q)\psi(\vec p_1) \rangle_0 \nonumber \\
&\tabi \times\Gamma^{ij}(\vec p_1,-\vec q,0).
\end{align}
The factor $i\eta$ with $\eta\rightarrow0^+$ is there to make sure that $\sigma^d$ is a retarded response function. Taking the thermodynamic limit, we have
\beq 
\lim\limits_{q\rightarrow0} \Gamma(\vec p,-\vec q,0) &=& 2\delta^{ij}K'(\vec p^2) + 4p^ip^jK''(\vec p^2) \nonumber \\
&=& \frac{\partial^2}{\partial p_i \partial p_j}K(\vec p^2).
\eeq
Finally, the diamagnetic conductivity is given by
\beq \label{eq:sigma_d}
\sigma^d_{ij}(\omega) &=& \lim\limits_{q\rightarrow 0}\sigma^d_{ij}(\vec q,\omega) \nonumber \\ 
&=& \frac{ie^2}{\omega + i\eta}\int\frac{d^dp}{(2\pi)^{d}}n(\vec p)\frac{\partial^2}{\partial p_i \partial p_j}K(\vec p^2)
\eeq
where $n(\vec p) \equiv \langle\psi^\dagger(\vec p)\psi(\vec p)\rangle_0$ is an occupation number of the momentum state $\vec p$. 

We now calculate the paramagnetic conductivity from $\delta\langle J_i \rangle$. We can drop the terms with $A_j$ in $J_i$ (the second term in Eq. (\ref{eq:J_i})) inside the commutator, since they contribute to a non-linear response. From the assumption of a uniform background field, we have $A_j(\vec x,t) = A_j(t)$ in Eq. (\ref{eq:lin_response}). Performing the Fourier transform on $\delta \langle J_i \rangle$ and then taking the thermodynamic limit, one obtains
\beq
\delta \langle J_i \rangle(t) = -i \int\limits_{-\infty}^{t} dt' \langle [\tilde{J}_i(t),\tilde{J}_j(t')] \rangle_0 A_j(t'),
\eeq
where $\tilde{J}_i(t) \equiv \int d^dx J_i(\vec x,t)$. We define the response function as $\chi_{ij}(t,t') \equiv -i\Theta(t-t')\langle[ \tilde{J}_i(t),\tilde{J}_j(t') ]\rangle_0$. As a result of  time-translational invariance of the system, $\chi_{ij}(t,t') = \chi_{ij}(t-t') = -i\Theta(t-t')\langle[ \tilde{J}_i(t-t'),\tilde{J}_j(0)]\rangle_0$. As a result, we find $\delta \langle J_i \rangle$ in frequency space is given by
\beq \label{eq:delta_Ji}
\delta \langle J_i \rangle(\omega) = \chi_{ij}(\omega) A_j(\omega),
\eeq
with
\begin{align} \label{eq:chi_sum}
\chi_{ij}(\omega) = \sum\limits_{m\neq n} \frac{e^{-\beta E_n}}{Z} \bigg(&\frac{\langle \psi_n | \tilde{J_i} | \psi_m \rangle \langle \psi_m | \tilde{J_j} | \psi_n \rangle}{\omega - (E_m - E_n) + i\eta} \nonumber \\
& - \frac{\langle \psi_n | \tilde{J_j} | \psi_m \rangle \langle \psi_m | \tilde{J_i} | \psi_n \rangle }{\omega - (E_n - E_m) + i\eta} \bigg).
\end{align}
Here $\tilde{J} \equiv \tilde{J}(t=0)$ and the summation in Eq. (\ref{eq:chi_sum}) is over all eigenstates of $H$ from Eq. (\ref{eq:H}). Using Eq. (\ref{eq:delta_Ji}), we rewrite the paramagnetic conductivity as
\beq \label{eq:sigma_p}
\sigma^p_{ij}(\omega) = \frac{i}{\omega+i\eta}\chi_{ij}(\omega).
\eeq
Combining the results from Eqs. (\ref{eq:sigma_d}) and (\ref{eq:sigma_p}), we finally obtain the total conductivity
\begin{align}
\sigma_{ij}(\omega) =& \frac{ie^2}{\omega + i\eta}\int\frac{d^dp}{(2\pi)^{d}}n(\vec p)\frac{\partial^2}{\partial p_i \partial p_j}K(\vec p^2) \nonumber \\
&+ \frac{i}{\omega+i\eta}\chi_{ij}(\omega).
\end{align}
To derive the sum rule for the $ii$ component of the optical conductivity, we utilize the Kramers-Kronig relation,
\beq \label{eq:Kramers}
\sigma_{2}(\omega) = -\frac{1}{\pi}\int\limits_{-\infty}^{\infty} d\omega' P\frac{\sigma_{1}(\omega')}{\omega' - \omega} ,
\eeq
where $\sigma_{1}$ and $\sigma_{2}$ denote the real part and the imaginary parts of $\sigma_{ii}$, respectively. $P$ denotes the Cauchy principal integral. Taking the limit $\omega \rightarrow \infty$ in Eq. (\ref{eq:Kramers}), one finds $\int_{-\infty}^{\infty}\sigma_{1}(\omega)d\omega =\pi \lim\limits_{\omega \rightarrow \infty}\omega\sigma_{2}(\omega)$.
Using the fact that $\sigma_1$ is even, we obtain the sum rule
\beq
W \equiv \int\limits_{0}^{\infty} \sigma_1(\omega) d\omega = \frac{\pi e^2}{2} \int \frac{d^dp}{(2\pi)^d} n(\vec p)\frac{\partial^2}{\partial p_i^2}K(\vec p^2). \label{eq:sum_rule}
\eeq 
We can neglect the paramagnetic part when taking the limit $\lim\limits_{\omega\rightarrow \infty}\omega\sigma_2(\omega)$ because $\sigma^d \sim \omega^{-1}$ and $\sigma^p \sim \omega^{-2}$ as $\omega\rightarrow\infty$.  The result coincides with the conductivity sum rule of particles in a restricted band (Eq. (\ref{eq:sumrule_restrictedband})). For the trivial case in which the kinetic term has a canonical form $K(\vec p^2) = \frac{p^2}{2m}$, the sum rule of $\sigma_{1}$ is given by $W = \frac{\pi e^2 n}{2m}$ as expected.

\section{Non-interacting Fermions} \label{sec:nonint_fermions}
We apply the conductivity sum rule derived above to a system of non-interacting fermions with the kinetic term of a form
\beq \label{eq:energy}
K(\vec p^2) = cp^{2\alpha},
\eeq
where $c$ and $\alpha$ are positive real constants. The constant $c$ has units of $[E]^{1-2\alpha}$ where $[E]$ denotes units of energy. The potential of this system is assumed to be weak enough such that the low energy (or small momentum) behavior of the total energy is the same as the kinetic term.\footnote{It is possible that, due to the potential, the constant $c$ is renormalized to be $c'$. However, using $c$ instead of $c'$ in $\varepsilon_{\vec p}$ will not change the powers of $n$ and $T$ we obtain in the sum rule. So, for simplicity, we will use $c$ in our calculation.} That is, the total energy $\varepsilon_{\vec p} = K(\vec p^2) =  cp^{2\alpha}$ when $p$ is less than a large momentum cutoff $\Lambda$. For simplicity, we will take $\varepsilon_{\vec p} =  cp^{2\alpha}$ for the whole range of $p$. This approximation is valid as long as $T \ll \varepsilon_\Lambda$. Since this is a non-interacting-fermionic system, the occupation number of the momentum state $\vec p$ is given by the Fermi-Dirac distribution,
\beq \label{eq:fermi_dirac}
n(\vec p) = \frac{1}{e^{\beta (\varepsilon_{\vec p}-\mu)} + 1},
\eeq 
where $\mu$ is the chemical potential. The density is the integral of $n(\vec p)$ over all momenta,
\beq
n = \int \frac{d^dp}{(2\pi)^d} n(\vec p). \label{eq:density}
\eeq
We calculate the sum rule of this system in the large (Appendix \ref{app:highT}) and low temperature limits (Appendix \ref{app:lowT}). The result is
\beq
\frac{W}{\pi e^2} \approx  
\begin{cases} \label{eq:sum_rule_expansions}
D c^{\frac{1}{\alpha}} n T^{\frac{\alpha-1}{\alpha}}  \ \ \ \ \ \text{if   } n \ll (\frac{T}{c})^{\frac{d}{2\alpha}} \\
Acn^{1+\frac{2(\alpha-1)}{d}}  \ \ \ \  \text{if   } n \gg (\frac{T}{c})^{\frac{d}{2\alpha}} 
\end{cases}
\eeq 
where the constants $D = (\alpha + \frac{2\alpha(\alpha-1)}{d})\frac{\Gamma(\frac{d-2}{2\alpha}+1)}{\Gamma(\frac{d}{2\alpha})}$ and $A = \alpha(2\pi)^{2(\alpha-1)}(\frac{d}{S_d})^{\frac{2(\alpha-1)}{d}}$. We note that when $\alpha = 1$ and $c = \frac{1}{2m}$, we recover the standard result, $W = \frac{\pi e^2 n}{2m}$, in both limits. 

\begin{figure}[h!]
\centering
	\subfigure[$\ \alpha = 1/3$ \label{fig:wn1}]{\includegraphics[scale=0.5]{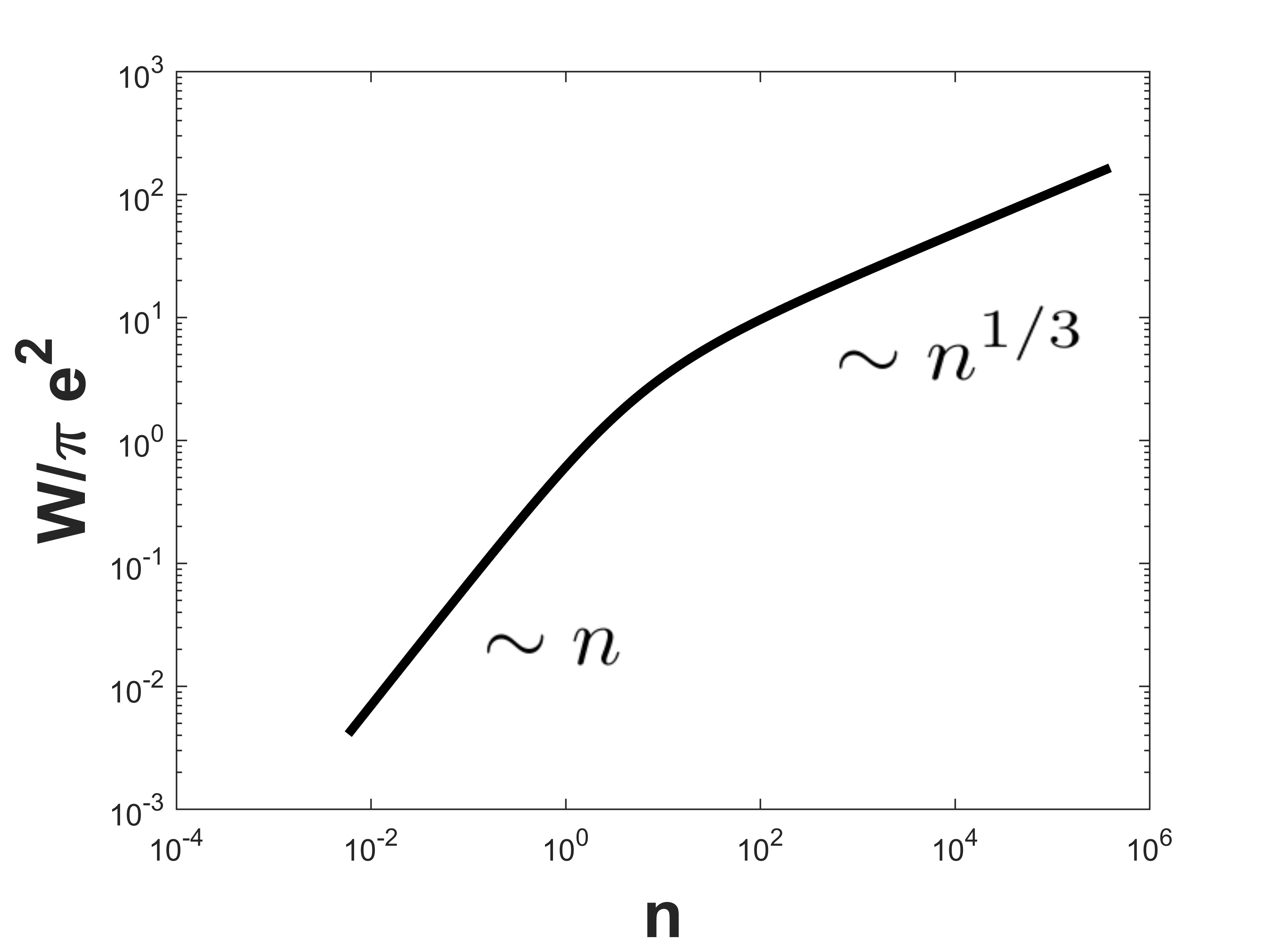}}
	\subfigure[$\ \alpha = 5/3$ \label{fig:wn2}]{\includegraphics[scale=0.5]{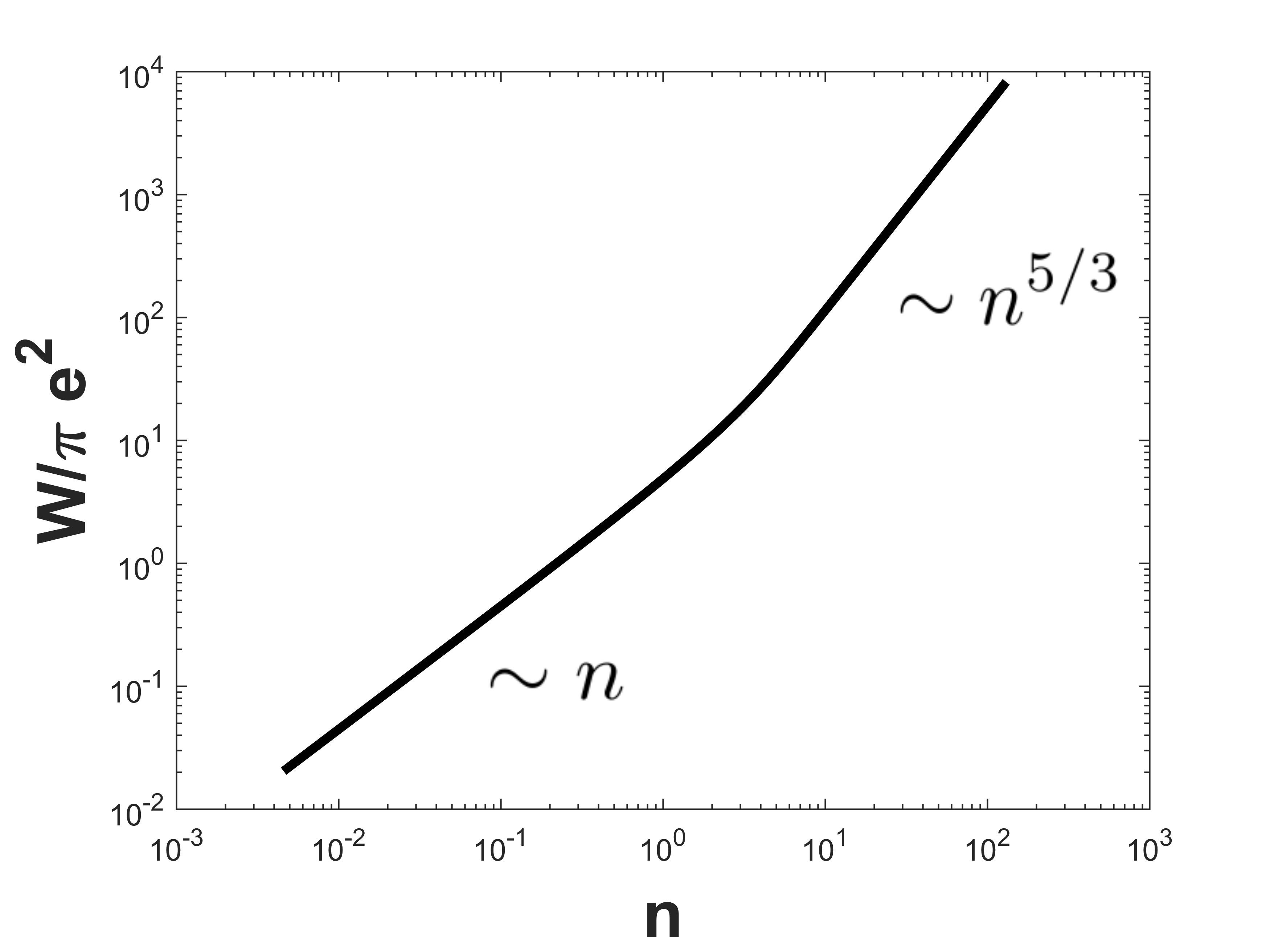}}
	\caption{Log-log plots of optical sum ($W$) vs. particle density ($n$) at two values of $\alpha$. We work in the units that $c = 1$. The parameters we use are $d = 2$ and $T = 0.5$.}
\end{figure}

We numerically evaluate the conductivity sum rule (Eq. (\ref{eq:sum_rule})). We display the results for the cases of $\alpha = 1/3$ in Fig. \ref{fig:wn1} and $\alpha = 5/3$ in Fig. \ref{fig:wn2}.
The numerical results confirm that $W$ has different behaviors at low densities and high densities for both $\alpha<1$ and $\alpha>1$ cases. 

\begin{figure}[h!]
	\includegraphics[scale=0.5]{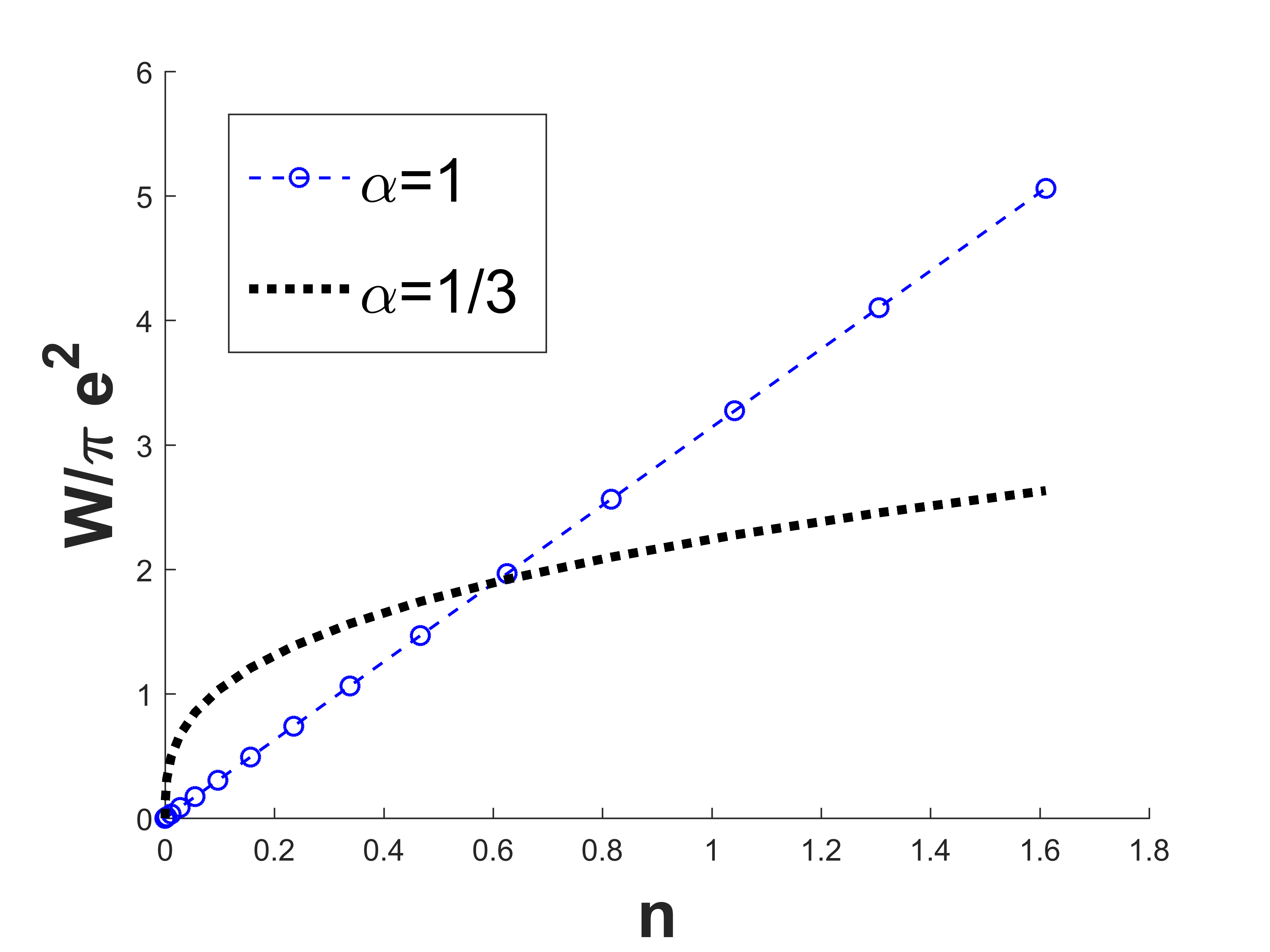} 
    \caption{Plots of optical sum ($W$) vs. particle density ($n$) for the cases of $\alpha = \frac{1}{3}$ and $1$. The parameters that we use are $T = 0.01$, $d = 2$. We set $c = 1$ for both $\alpha = 1$ and $\alpha = \frac{1}{3}$ cases.} \label{fig:sumrule}
\end{figure}

Using the result we obtain in this section, we can qualitatively explain the behavior  of the effective number of charge carriers, $N_{\rm eff}$, at various doping levels in the cuprates\cite{Cooper1990,Uchida1991}. When $0<x<0.2$, $N_{\rm eff}(x) \propto x^\gamma$ with $\gamma \approx 0.3 - 0.4$ as we have discussed in the introduction. Qualitatively matching this feature of $N_{\rm eff}$ with our model necessitates low temperatures and $0<\alpha<1$, and hence one has $W \propto n^{\beta} \propto x^\beta$ with $0<\beta = 1 + \frac{2(\alpha-1)}{d}< 1$. Here, as mentioned in the introduction, we make an assumption that the number of excitations with fractional kinetic energy is the same as that of mobile electrons or holes, $n \propto x$. As a concrete example, we make a plot of $W$ vs. $n$ in this low temperature limit with the exponent between $0$ and $1$ (for $\alpha = 1/3)$ in Fig. \ref{fig:sumrule}. The plot in the case of $\alpha = 1$ is also displayed for comparison. The region of $n$ for which $W(\alpha = 1/3) > W(\alpha = 0)$ has qualitatively the same feature as $N_{\rm eff}$ in the cuprates.  We note that there is no unit cell in the model we are using. This means we cannot numerically relate $W$ to $N_{\rm eff}$ and $n$ to $x$. As a result, rather than making a plot of $N_{\rm eff}$ against $x$ as in Refs. \cite{Cooper1990,Uchida1991}, we are restricted to the plot of $W$ vs. $n$. 

\section{Discussion and Conclusion}

The key result of this paper is that the conductivity sum rule of non-interacting fermions with a fractional kinetic energy does not follow the traditional result. At high temperatures and low densities, the optical sum scales as $W \propto nT^{\frac{\alpha-1}{\alpha}}$.  At low temperatures and high densities, the optical sum is given by $W \propto n^{1+\frac{2(\alpha-1)}{d}}$. One can use the result at low temperatures to qualitatively explain the behavior of $N_{\rm eff}$ at various doping concentration in the cuprates. To nail down that the current-carrying excitations in the cuprates are in fact governed by a fractional kinetic energy requires further experiments. That is, one needs to experimentally verify that the optical sum has two regimes as we have predicted in Eq. (\ref{eq:sum_rule_expansions}). This can be achieved by measuring the optical conductivity and then computing the empirical optical sum as a function of $x$ at higher temperatures. However, we must keep in mind that the temperature cannot be raised  too high because the assumption that the excitation energy, $\varepsilon_{\vec p}$ has the same form as the kinetic energy, $K(\vec p^2)$, will break down eventually. The assumption is valid only when $T\ll\varepsilon_\Lambda$. \\

\noindent \textbf{Acknowledgements} We thank the NSF DMR-1461952 for partial funding of this project. KL is supported by the Department of Physics at the University of Illinois and a scholarship from the Ministry of Science and Technology, Royal Thai Government. 

\onecolumngrid

\appendix

\section{Bloch Theorem for Non-canonical Kinetic Term} \label{app:bloch}

In this section, we show that the spontaneous current term in Eq. (\ref{eq:J_i_expect0}),
\beq
\int\frac{d^dp}{(2\pi)^{d}}\langle \psi^{\dagger}(\vec p)\psi(\vec p) \rangle_0 2p^iK'(\vec p^2) = 0,
\eeq
is zero in the thermodynamic limit.  We note that in Eq. (\ref{eq:J_i_expect0}), $\lim\limits_{\vec q \rightarrow 0}\Gamma^i(\vec p, -\vec q) = 2p^iK'(\vec p^2)$ . Our proof is based on Refs. \cite{Bohm1949,Ohashi1996}.

Let us introduce the momentum translation operator,
\beq
T(\vec p) \equiv e^{-i\vec p \cdot \vec R},
\eeq
where the operator $\vec R$ is defined as $\vec R \equiv \int d^dr \psi^\dagger (\vec r)\vec r \psi(\vec r)$. For small $\vec p'$, one can show that 
\beq \label{eq:p_trans}
T^{\dagger}(\vec p')\psi(\vec p)T(\vec p') &\approx&  \psi(\vec p) + i\vec p'\cdot[\vec R,\psi(\vec p)] \nonumber \\
&=& \psi(\vec p) - i\int d^dr e^{-i\vec p\cdot\vec r}\vec p'\cdot \vec r \psi(\vec r) \nonumber \\
&=& \psi(\vec p) + \vec p' \cdot \nabla_p \psi(\vec p) \nonumber \\
&\approx& \psi(\vec p + \vec p').
\eeq
On the first line, we use the identity $[\psi^\dagger(\vec r)\psi(\vec r),\psi(\vec r')] = -\delta^d(\vec r - \vec r')\psi(\vec r)$ which is valid for both fermionic and bosonic fields. In the same manner as Eq. (\ref{eq:p_trans}), one can show that $T^{\dagger}(\vec p')\psi^\dagger(\vec p)T(\vec p') = \psi^\dagger(\vec p + \vec p')$. 

Let $\{|\psi_i\rangle \}$ be a complete, orthonormal set of eigenstates and let the eigenenergy of the eigenstate $|\psi_i\rangle$ be $E_i$. We define the thermal equilibrium density matrix which gives the lowest free energy at temperature $T$ as
\beq
\rho_\psi \equiv \sum\limits_{i} |\psi_i\rangle w_i \langle \psi_i |,
\eeq
where $w_i = \frac{e^{-\beta E_i}}{\mathrm{Tr}(e^{-\beta H})}$ is a Boltzmann weight. The expectation $\langle O \rangle_0$ of an operator $O$ defined in the main text corresponds to $\mathrm{Tr}(\rho_\psi O)$. We assume that the expectation value of the current,
\beq \label{eq:sponJ}
J^i_\psi = \int\frac{d^dp}{(2\pi)^{d}}\mathrm{Tr}\bigg(\rho_\psi \psi^{\dagger}(\vec p)\psi(\vec p) \bigg)2p^iK'(\vec p^2) \neq 0.
\eeq 
with respect to $\rho_\psi$ is finite.  We show, in this appendix, that this assumption will lead to a contradiction. We introduce a trial density matrix,
\beq
\rho_\phi \equiv  \sum\limits_{i} |\phi_i\rangle w_i \langle \phi_i |.
\eeq
Here $\{ |\phi_i\rangle \}$ is another set of complete, orthonormal eigenstates defined by
\beq
|\phi_i\rangle \equiv T(-\delta\vec p)|\psi_i\rangle
\eeq
where $\delta \vec p$ is a small momentum parameter. Since, by construction, $\rho_\psi$ and $\rho_\phi$ have the same statistical weight, $w_i$, their entropies are equal: $S_\psi = S_\phi = -\mathrm{Tr}(\rho \ln \rho) = - \sum_{i}w_i \ln w_i $. The expectation value of the energy with respect to $\rho_\phi$ is
\beq \label{eq:E_phi}
E_\phi &=& \mathrm{Tr}(\rho_\phi H) = \sum\limits_{i} w_i\langle \phi_i | H | \phi_i \rangle \nonumber \\
&=& \sum\limits_{i} w_i\langle \psi_i |T^\dagger(-\delta \vec p) H T(-\delta \vec p)| \psi_i \rangle \nonumber \\
&=& \mathrm{Tr} \bigg(\rho_\psi T^\dagger(-\delta \vec p) H T(-\delta \vec p) \bigg).
\eeq
For the kinetic part of the Hamiltonian, $H_K = \int \frac{d^dp}{(2p)^d}\psi^\dagger(\vec p)\psi(\vec p)K(\vec p^2)$, we find that
\beq
T^\dagger(-\delta \vec p) H_K T(-\delta \vec p) &=&  \int \frac{d^dp}{(2p)^d}T^\dagger(-\delta \vec p)\psi^\dagger(\vec p)T(-\delta \vec p) T^\dagger(-\delta \vec p)\psi(\vec p)T(-\delta \vec p) K(\vec p^2) \nonumber \\
&=& \int \frac{d^dp}{(2p)^d}\psi^\dagger(\vec p - \delta \vec p)\psi(\vec p - \delta \vec p) K(\vec p^2) \nonumber \\
&=& \int \frac{d^dp}{(2p)^d}\psi^\dagger(\vec p)\psi(\vec p) K((\vec p + \delta\vec p)^2) \nonumber \\
&\approx& H_K + \delta\vec p\cdot\int \frac{d^dp}{(2p)^d}\psi^\dagger(\vec p)\psi(\vec p)2\vec p K'(\vec p^2) + O(\delta\vec p^2).
\eeq
On the first line, we use Eq. (\ref{eq:p_trans}) and its complex conjugate to translate the momentum of the field operators by $-\delta\vec p$. Because there is no derivative terms in other parts of the Hamiltonian, the momentum translation leaves them invariant. As a result, one finds
\beq \label{eq:trans_H}
T^\dagger(-\delta \vec p) H T(-\delta \vec p) &=& H + \delta\vec p\cdot\int \frac{d^dp}{(2p)^d}\psi^\dagger(\vec p)\psi(\vec p)2\vec p K'(\vec p^2) + O(\delta\vec p^2).
\eeq 
Using Eqs. (\ref{eq:sponJ}), (\ref{eq:E_phi}), and (\ref{eq:trans_H}), we rewrite the energy of $\rho_\phi$ as
\beq
E_\phi &=& \mathrm{Tr}(\rho_\psi H ) + \delta\vec p\cdot\int \frac{d^dp}{(2p)^d}\mathrm{Tr}\bigg(\rho_\psi \psi^\dagger(\vec p)\psi(\vec p)\bigg)2\vec p K'(\vec p^2) \nonumber \\
&=& E_\psi + \delta\vec p\cdot \vec J_\psi.
\eeq
The free energy of $\rho_\phi$ is
\beq
F_\phi = E_\phi - TS_\phi = F_\psi +  \delta\vec p\cdot \vec J_\psi.
\eeq
If we choose $\delta \vec p$ to have the opposite direction as $\vec J_\psi$, we find $F_\phi < F_\psi$. This result contradicts the assumption that $\rho_\psi$ has the lowest free energy. Consequently, the spontaneous current $\vec J_\psi$ is zero.

\section{High Temperature Expansion} \label{app:highT}

We investigate the conductivity sum rule of non-interacting fermions at high temperatures and low densities. We first perform a high temperature expansion on the Fermi-Dirac distribution to obtain the fugacity as a function of density and temperature \cite{Huang1987}. We rewrite Eq. (\ref{eq:density}) as
\beq
n\lambda^d = \frac{2\alpha}{\Gamma(\frac{d}{2\alpha})}\int\limits_{0}^{\infty}\frac{x^{d-1}}{z^{-1}e^{x^{2\alpha}}+1}dx,
\eeq
where $z \equiv e^{\beta\mu}$ is the fugacity, $\lambda \equiv 2\pi(\frac{c}{T})^{\frac{1}{2\alpha}}(\frac{2\alpha}{S_d\Gamma(\frac{d}{2\alpha})})^{\frac{1}{d}}$ is the thermal de Broglie wavelength, and $S_d = \frac{2\pi^{\frac{d}{2}}}{\Gamma(\frac{d}{2})}$ is a surface area of a unit ($d-1$)-sphere. Expanding the right-hand-side in powers of $z$, one finds
\beq
n \lambda^d = \sum\limits_{n=0}^{\infty}\frac{(-1)^nz^{n+1}}{(n+1)^{\frac{d}{2\alpha}}}.
\eeq
We then solve for $z$ in term of $n\lambda^d$ by substituting $z = \sum\limits_{m = 1}^{\infty} a_m (n\lambda^d)^m$ and then matching the coefficients of $(n\lambda)^l$. The result is
\beq
z = n\lambda^d + \frac{1}{2^{\frac{d}{2\alpha}}}(n\lambda^d)^2 + O((n\lambda^d)^3).
\eeq
At high temperatures, one can omit the higher order term in $n\lambda^d$ and thus
\beq
z \approx n\lambda^d.
\eeq
It follows that $n(p)$ in the high $T$ limit is given by
\beq \label{eq:np_large_T}
n(p) = n\lambda^d e^{-\beta \varepsilon_{\vec p}}.
\eeq
Substituting Eq. (\ref{eq:np_large_T}) into Eq. (\ref{eq:sum_rule}) and then evaluating the momentum integral, we obtain the sum rule,
\beq 
\frac{W}{\pi e^2}&=&D c^{\frac{1}{\alpha}} n T^{\frac{\alpha-1}{\alpha}}
\eeq
where $D = (\alpha + \frac{2\alpha(\alpha-1)}{d})\frac{\Gamma(\frac{d-2}{2\alpha}+1)}{\Gamma(\frac{d}{2\alpha})}$ is a constant. This result is valid when $n\lambda^d \ll 1$ or $n \ll (\frac{T}{c})^{\frac{d}{2\alpha}}$.

\section{Low Temperature Expansion} \label{app:lowT}

We perform the Sommerfeld expansion\cite{Ashcroft1976} on Eq. (\ref{eq:sum_rule}) to investigate the low temperature ($T\ll\varepsilon_F$) and high density behavior of the conductivity sum rule for non-interacting fermions. Using equation $n = \int\limits_{p<p_F}d^dp$, one can relate the density, $n$, to Fermi momentum, $p_F$, as 
$p_F = 2\pi(\frac{d}{S_d})^{1/d} n^{1/d}$,
where $S_d = \frac{2\pi^{\frac{d}{2}}}{\Gamma(\frac{d}{2})}$ is a surface area of a unit ($d-1$)-sphere. From $\varepsilon_{\vec p} = cp^{2\alpha}$, one finds the Fermi energy is given by
\beq \label{eq:ef}
\varepsilon_F = c(2\pi)^{2\alpha}(\frac{d}{S_d})^{\frac{2\alpha}{d}}n^{\frac{2\alpha}{d}}.
\eeq
We solve Eq. (\ref{eq:density}) for $\mu$ using the Sommerfeld expansion\cite{Ashcroft1976}
\beq \label{eq:summerfeld}
\int\limits_{-\infty}^{\infty} d\varepsilon \frac{H(\varepsilon)}{e^{\beta(\varepsilon - \mu)}+1} &\approx& \int\limits_{0}^{\mu} H(\varepsilon) d\varepsilon + \frac{\pi^2}{6}H'(\mu)T^2 \nonumber \\
&\approx& \int\limits_{0}^{\varepsilon_F} H(\varepsilon) d\varepsilon + (\mu - \varepsilon_F)H(\varepsilon_F) + \frac{\pi^2}{6}H'(\varepsilon_F)T^2.
\eeq
The result is
\beq \label{eq:mu}
\mu = \varepsilon_F - \frac{\pi^2}{6}(\frac{d}{2\alpha}-1)\frac{T^2}{\varepsilon_F}. \label{eq:chemical_potential}
\eeq
In the next step, we use the Sommefeld expansion on Eq. (\ref{eq:sum_rule}). We substitute the chemical potential (Eq.(\ref{eq:mu})) and Fermi energy (Eq.(\ref{eq:ef})) into the resulting expansion. We are then able to rewrite the sum rule at low temperature as
\beq \label{eq:W_highT}
\frac{W}{\pi e^2} = Acn^{1+\frac{2(\alpha-1)}{d}} + B\frac{(\alpha-1)(d+2(\alpha-1))T^2}{c}n^{1-\frac{2(\alpha+1)}{d}}.
\eeq
$A$ and $B$ are positive constants given by $A = \alpha(2\pi)^{2(\alpha-1)}(\frac{d}{S_d})^{\frac{2(\alpha-1)}{d}}$ and $B = \frac{\pi^2}{12\alpha}\frac{1}{(2\pi)^{2(\alpha+1)}}(\frac{S_d}{d})^{\frac{2(\alpha+1)}{d}}$. This result is valid when $T \ll \varepsilon_F$ or $n \gg (\frac{T}{c})^{\frac{d}{2\alpha}}$.

\twocolumngrid

\bibliography{optsum}
\bibliographystyle{apsrev4-1}
\end{document}